# Phase and sulfur vacancy engineering in cadmium sulfide for boosting hydrogen production from catalytic plastic waste photoconversion

Thanh Tam Nguyen[a,b], Jacqueline Hidalgo-Jiménez[a,c], Xavier Sauvage[d], Katsuhiko Saito[e], Qixin Guo[e], and Kaveh Edalati[a,b,*]

[a] WPI, International Institute for Carbon Neutral Energy Research (WPI-I2CNER), Kyushu University, Fukuoka 819-0395, Japan
[b] Mitsui Chemicals, Inc. - Carbon Neutral Research Center (MCI-CNRC), Kyushu University, Fukuoka 819-0395, Japan
[c] Graduate School of Integrated Frontier Sciences, Department of Automotive Science, Kyushu University, Fukuoka 819-0395, Japan
[d] Univ Rouen Normandie, INSA Rouen Normandie, CNRS, Groupe de Physique des Matériaux, UMR6634, 76000 Rouen, France
[e] Department of Electrical and Electronic Engineering, Synchrotron Light Application Center, Saga University, Saga 840-8502, Japan

Cadmium sulfide (CdS) is a well-known low-bandgap photocatalyst, but its efficiency is often hindered by rapid photo-generated carrier recombination and a limited number of active catalytic sites. To overcome these challenges, this study introduces an efficient CdS photocatalyst through a novel strategy combining metastable-to-stable phase transformation and sulfur vacancy generation. This strategy integrates hydrothermal treatment and a high-pressure process to create sulfur vacancies, which serve as active catalytic sites, within a thermodynamically stable wurtzite (hexagonal) phase known for its superior photocatalytic properties. The resulting CdS photocatalyst demonstrates exceptional performance in photoreforming for hydrogen production and the conversion of polyethylene terephthalate (PET) plastic into valuable materials. Compared to commercial CdS catalysts, this new material shows a 23-fold increase in both hydrogen production and plastic degradation without the need for co-catalysts. Quenching experiments reveal that holes and hydroxyl radicals play crucial roles in the photoreforming process of this vacancy-rich CdS. First-principles calculations via density functional theory (DFT) indicate that the hexagonal phase possesses a lower bandgap and it exhibits further bandgap narrowing with the introduction of sulfur vacancies. These findings not only present an innovative approach to CdS processing but also highlight the critical role of sulfur vacancies as effective defects for the catalytic photoreforming of microplastics.

*Corresponding author: Kaveh Edalati (E-mail: kaveh.edalati@kyudai.jp; Tel: +80-92-802-6744)

## 1. Introduction

The contemporary world grapples with a myriad of crises, among which $CO_2$ emissions [1] and plastic waste pollution [2] stand out as particularly pressing environmental concerns. Across nations, efforts are underway to mitigate annual global greenhouse gas emissions by embracing $H_2$ as a clean fuel alternative devoid of $CO_2$ emissions [3,4]. However, the clean production of $H_2$ remains a formidable challenge. While $H_2$ holds promise as an energy carrier capable of facilitating a transition towards a net-zero carbon society, existing production technologies prove either costly or unsustainable. Notably, the prevalent method of steam reforming of fossil fuels accounts for the majority of the world's $H_2$ supply but contributes significantly to $CO_2$ emissions [5]. Alternatively, photocatalytic hydrogen production utilizing water and solar energy emerges as a viable avenue for generating $H_2$ sustainably. Nevertheless, this approach encounters obstacles such as low photocatalyst efficiency and the necessity of sacrificial agents.

Another global challenge is plastic pollution. The proliferation of plastic waste in the environment, resulting in the formation of microplastics or nanoplastics, poses a grave threat to ecosystems and human health. In response to these intertwined energy and environmental issues, the concept of photoreforming was introduced [6]. The photoreforming process entails the photodegradation of plastic waste, serving as sacrificial agents, to facilitate $H_2$ production from water. In the pursuit of efficient photoreforming, various low-cost options have been explored, including metal oxides [7-11], metal phosphides [7,9], metal sulfides [12-16] and organic materials [17]. Metal sulfides, particularly cadmium sulfide (CdS), have garnered significant attention due to their favorable photochemical properties and quantum efficiency [18,19].

CdS has been extensively utilized in various applications including infrared applications, photovoltaics, sensors, energy storage applications, and photocatalysis. In the photocatalytic field, it has been used in photocatalytic hydrogen production, reduction of carbon dioxide, and degradation of pollutants [20-22]. However, the intrinsic drawbacks of CdS include photo-corrosion, undesirable charge carrier recombination, and limited active sites [20,21]. To address these challenges, researchers have endeavored to design novel photocatalysts by manipulating morphology [23], constructing heterojunction structures [24,25], and doping [26]. It is well known that CdS has two main structural polytypes, wurtzite and zinc blende structures, belonging to hexagonal and cubic crystal systems, respectively, as shown in Fig. 1a [27]. While hexagonal CdS exhibits the highest thermodynamic stability, cubic CdS is a metastable phase that undesirably forms. Maintaining CdS in the hexagonal structure is crucial for practical applications due to its superior optoelectronic properties. In the hexagonal phase, the distortion of the $CdS_4$ tetrahedron engenders an internal electric field, facilitating the efficient photo-generated electron-hole separation, an essential aspect of photocatalytic reactions. Conversely, cubic CdS lacks this internal electric field, leading to slower photo-generated carrier migration rates [19]. Consequently, stabilization of hexagonal CdS is a key solution to achieve a higher photocatalytic activity. Additionally, although the effect of vacancies on the photoreforming of CdS has received little attention, vacancies are expected to modify the electron band structure, phonon vibration, and photo-generated carrier concentration. Vacancies in the crystal structure can also act as electron trap and adsorption centers, facilitating carrier spatial separation and promoting efficient charge transfer routes, thus augmenting photocatalytic performance [28-31]. Unlike oxygen vacancies, there have been limited studies on the significance of sulfur vacancies on photocatalysis. In a study, sulfur vacancies were introduced to successfully capture photo-generated electrons and extend the charge carrier lifetime in the monolayer $ZnIn_2S_4$ structure [32]. In another study, CdZnS solid solution rich in sulfur vacancies was developed and showed that sulfur vacancies acted as adsorption centers for hole consumption, subsequently leading to a high photocatalytic $H_2$ production [29]. Taken together, the introduction of sulfur vacancies in hexagonal CdS can be a potential solution to enhance its photocatalytic activity for photoreforming.

In this study, we present a novel approach to enhance the photoreforming performance of CdS for simultaneous polyethylene terephthalate (PET) plastic waste conversion and $H_2$ production without co-catalyst addition. While earlier attempts focused on the modification of CdS catalysts by introducing dopants or composite heterojunctions [23-25], this study develops a pure form of CdS rich in hexagonal phase and sulfur vacancies using an innovative two-step method of hydrothermal (schematically shown in Fig. 1b) and high-pressure torsion (HPT) (Fig. 1c). The superior catalytic performance of the synthesized material not only highlights the significance of sulfur vacancies and hexagonal phase in photoreforming but also underscores the potential of coupled hydrothermal and HPT approaches in developing active catalysts for sustainable energy and waste management.

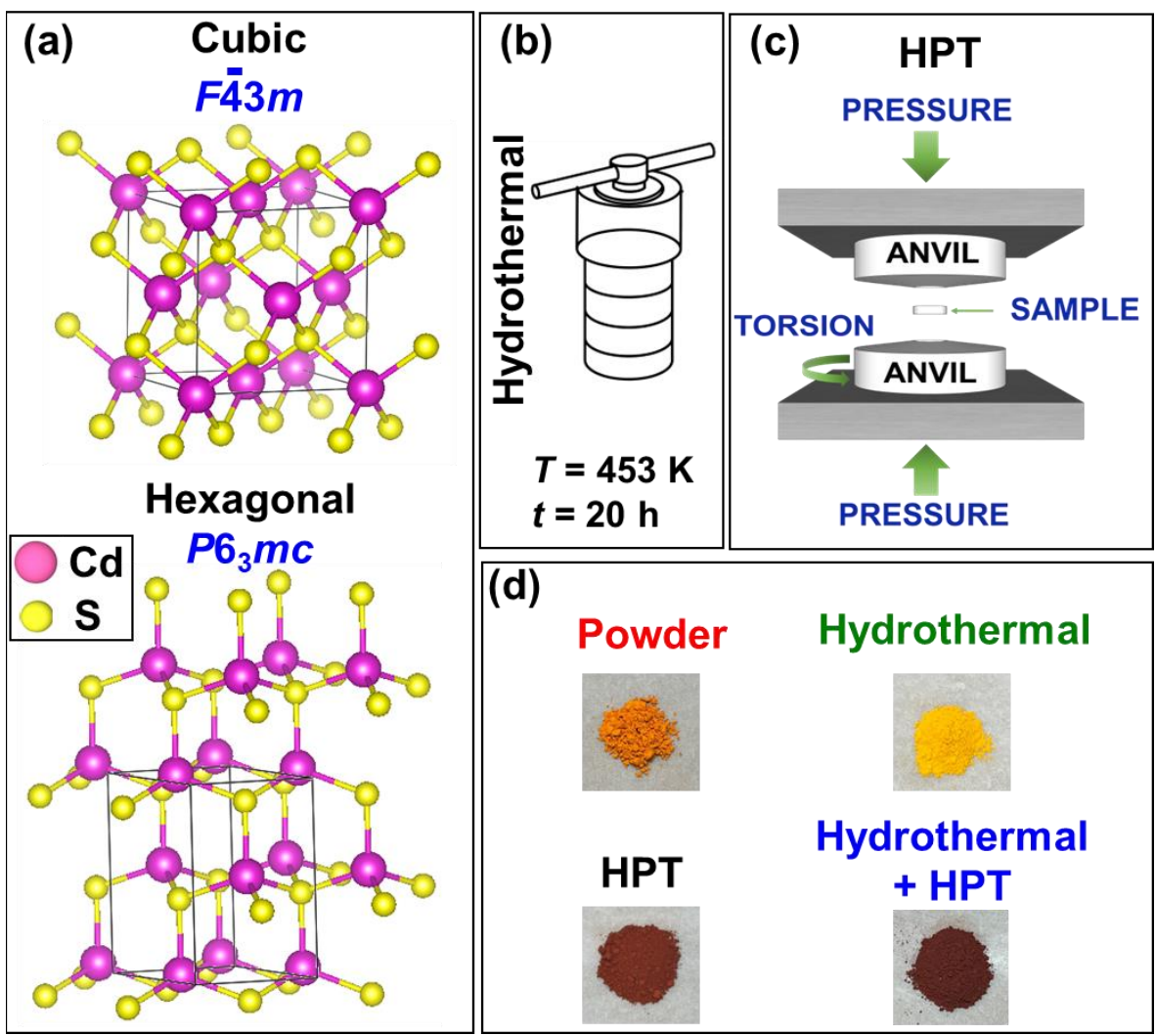

Figure 1. (a) Crystal images of cubic and hexagonal CdS phases, (b) schematic of hydrothermal process, (c) schematic of HPT process, and (d) appearance of samples.

## 2. Materials and methods
### 2.1. Materials

CdS powder was from Sigma-Aldrich, USA. NaOH was purchased from Fujifilm, Japan and a solution of 10 M NaOD (40 wt% in $D_2O$ with a purity of 99 at%) was also obtained from Sigma-Aldrich, USA. Microplastic PET powder with particle sizes less than 300 μm was purchased from GoodFellow Company, UK. Coumarin and maleic acid were obtained from Tokyo Chemical Industry Co., LTD and Nacalai Tesque, Inc., Japan, respectively. Hydrogen hexachloroplatinate (IV) hexahydrate ($H_2PtCl_6 \cdot 6H_2O$) was from Sigma-Aldrich, USA. Ethylenedinitrilotetracetic acid disodium salt dihydrate (EDTA-2Na) and methanol ($CH_3OH$) were obtained from Merck, USA.

Isopropanol (IPA) and 1,4-benzoquinone (BQ) were from Sigma-Aldrich, USA. All chemicals were utilized without any purification process.

## 2.2. Catalyst synthesis

The catalyst was synthesized using a two-step process involving hydrothermal treatment and HPT. Initially, 1 g of CdS powder was mixed with 100 mL of 10 M NaOH solution and subjected to an autoclave for hydrothermal treatment at 453 K for 20 h. The hydrothermal temperature and time were selected to have the maximum fraction of the hexagonal phase. Then, the mixture was cooled and washed with deionized water, and the CdS was collected by filtration. The NaOH solution remaining after the hydrothermal treatment was preserved for subsequent photocatalytic processes. Next, 300 mg of the hydrothermally treated CdS was compressed into a disc with a diameter of 10 mm under 0.38 GPa pressure. This disc was placed in a 10 mm diameter and 0.25 mm deep hole of a lower anvil of an HPT machine. The lower anvil was gradually raised to touch the anvil on the top and then rotated at a speed of 1 rpm under 6 GPa for 3 turns. The high pressure of 6 GPa was utilized because high pressure is an important factor enabling the deformation of hard ceramics like CdS [33]. The rotation speed of 1 rpm, which is a slow rate was intentionally selected to avoid the temperature rise during HPT [34]. The number of turns was selected as 3 because earlier studies on some ceramics suggested 2-4 turns are good enough to achieve desirable microstructural modifications [35-37]. This catalyst treated by hydrothermal followed by HPT is denoted as the Hydrothermal+HPT sample and used for the next experiments. Together with the hydrothermal+HPT sample, the samples of CdS initial powder, hydrothermal-processed powder and HPT-processed powder without hydrothermal treatment were also used for comparison. The sample images are shown in Fig. 1d, indicating that the sample color turns darker after HPT processing due to the generation of color centers such as vacancy-type defects.

## 2.3. Catalyst characterization

The crystalline structure of the four catalysts (powder, hydrothermal, HPT, hydrothermal+HPT) was evaluated using X-ray diffraction (XRD) with Cu K$\alpha$ irradiation having a wavelength of $\lambda = 0.1542$ nm. The Rietveld refinement using a PDXL software was used to analyze the fraction of phases and their lattice parameters. The crystallite size was calculated through the Halder-Wagner method [38]. Raman analysis was conducted by a $\lambda = 532$ nm laser light to further evaluate the crystalline structures. The microstructure of catalysts was examined by scanning electron microscopy (SEM) at a voltage of 5 kV and transmission electron microscopy (TEM) at a voltage of 200 kV. For TEM, small sample quantities were crushed in ethyl alcohol and distributed on carbon grids. TEM analyses included the bright- and dark-field (BF and DF) images, selected area electron diffraction (SAED), and high-resolution (HR) mode followed by analysis through fast Fourier transform (FFT). The surface areas of catalysts were measured by $N_2$ gas adsorption via the Brunauer-Emmett-Teller (BET) technique. X-ray photoelectron spectroscopy (XPS) using Al-Kα as an irradiation source was performed to analyze the surface electronic state of four catalysts, with the C1s core level at 284.8 eV as an internal reference for peak position correction. Electron paramagnetic resonance (EPR) analysis, utilizing a microwave with a frequency of 9.4688 GHz source at room temperature, was performed to detect sulfur vacancies. Ultraviolet-visible diffuse reflectance (UV-vis) spectroscopy was used to measure light absorbance, while the bandgap was calculated using the Kubelka-Munk method. XPS data was used to estimate the valence band top, while the conduction band bottom was estimated from the valence band and bandgap. Radiative electron-hole recombinations were examined using photoluminescence (PL) spectroscopy at ambient temperature using a laser light source with an excitation wavelength of 325 nm.

## 2.4. Photoreforming of plastic

For photocatalytic hydrogen production and plastic degradation, 50 mg of CdS was mixed with 50 mg of PET and added to 3 mL 10 M NaOH solution. For comparison, $Pt(NH_3)_4(NO_3)_2$ was added to some solutions to provide 1 wt% platinum as a co-catalyst. Then, the mixture was sonicated for 5 minutes and then purged with argon for 30 minutes. Following air evacuation, the mixture was exposed to 300 W Xenon light irradiation, with constant stirring at 298 K. Hydrogen production was quantified using gas chromatography having a thermal conductivity detector (GC-TCD). Plastic photodegradation components were recognized using $^1$H nuclear magnetic resonance (NMR) spectroscopy in a 10 M NaOD solution in $D_2O$. Maleic acid served as an internal standard for quantitative analysis of the photo-oxidized products by $^1$H NMR [39]. Moreover, the main reactive species were determined by quenching experiments with the addition of different scavengers to the photocatalytic system. The scavengers included IPA (0.02 mmol/L) for trapping hydroxyl (•OH) radicals and EDTA-2Na and $CH_3OH$ (10 vol%) for trapping photogenerated holes. To evaluate the extent of •OH radicals formed during photocatalysis, 1 mM coumarin was added to the photocatalytic reactor to form 7-hydroxycoumarin and the solution was subsequently evaluated by a spectrofluorometer utilizing a 332 nm excitation wavelength by monitoring the peal located at 450 nm [40].

## 2.5. First-Principles Calculations

To understand the behavior of CdS for photocatalysis, two points were theoretically investigated: (i) the bandgap of the hexagonal phase and (ii) the role of sulfur vacancies in the band structure. First-principles calculations were carried out using density functional theory (DFT). All procedures were performed within generalized gradient approximation (GGA) together with Perdew-Burke-Ernzerhof (PBE) [41] exchange-correlation functional implemented in Vienna *Ab initio* Simulation Package (VASP) [42,43]. Projected-augmented wave (PAW) pseudopotential described 12 valence electrons for Cd ($4s^2\ 5d^{10}$) and 6 electrons for S ($3s^2\ 3p^4$). The electronic wave function was expanded with a cutoff energy of 500 eV. The Brillouin zone was sampled with an optimized *K*-point mesh according to the geometric nature of each structure. Since DFT tends to underestimate the bandgap, the Hubbard *U* parameter was utilized which allows correction of the interaction between strongly correlated electronic states, most likely *d* or *f* orbitals. The comparison of the results with different *U*-parameters was performed for both cubic and hexagonal unit cells.

To investigate the significance of sulfur vacancies, supercells of the cubic and the hexagonal phases were generated with 64 atoms (32 Cd atoms and 32 S atoms). For this study, two scenarios were analyzed: (i) 3 at% sulfur vacancies by removing one sulfur atom from the pristine supercell, and (ii) 6 at% sulfur vacancies by removing two sulfur atoms. All the supercells were optimized before the self-consist field (SCF) and density of states (DOS) calculations. The bond length around sulfur vacancies were calculated and compared with those of the pristine structure. Moreover, the vacancy formation energy ($E_{vac}$) was calculated using the following relationship.

$E_{vac} = [(E_{def} + n\mu_S) - E_{pristine}] / n$                                                               (1)

where $E_{def}$ is the total energy of the structure with a defect, $E_{pristine}$ is the total energy of the structure without a defect, *n* represents the number of vacancies, and $\mu_S$ is the chemical potential for a single sulfur atom.

## 3. Results

### 3.1. Catalyst characterization

The phase structures of four CdS catalysts including powder, hydrothermal, HPT, and hydrothermal+HPT were evaluated using XRD and Raman spectroscopy, as illustrated in Fig. 2. Four CdS catalysts possess hexagonal and cubic phases in their structure. As illustrated in Table 1, the initial CdS powder has 46.3 wt% of the cubic phase and 53.7 wt% of the hexagonal phase. Under the hydrothermal process, the catalyst turns to the hexagonal structure with a 99.2 wt% fraction, indicating the successful phase transformation from cubic to the desirable hexagonal phase by the hydrothermal process. The CdS transformation from the cubic phase to the hexagonal phase is accompanied by an increase in crystallite size from 3.2 nm to 40.7 nm (Table 1) [19]. Conversely, HPT processing of initial powders leads to an undesirable phase transformation from hexagonal to cubic, resulting in 97.4 wt% of the cubic phase and 2.6 wt% of the hexagonal phase. The application of HPT to the sample processed by hydrothermal also leads to the formation of the cubic phase, but the catalyst still possesses 74.0 wt% of the hexagonal phase which is notably higher than that of the initial CdS powder. Therefore, although neither the hydrothermal method produces 100% of the hexagonal phase, nor the HPT method produces 100% of the cubic phase, the combination of these two synthesis methods can successfully transform CdS into a hexagonal-rich catalyst, which is a more stable and favorable for photoreforming [27,40].

The driving force for the cubic-to-hexagonal phase transformation by hydrothermal treatment is the lower Gibbs free energy of the thermodynamically stable hexagonal phase [27]. Although this phase transformation should thermodynamically occur even at ambient temperature, the transition is suppressed by a high energy barrier (i.e. kinetics activation energy). Another reason for the stability of the cubic CdS phase is that the driving force for a metastable-to-stable phase transition is lower for nanocrystalline materials due to the contribution of grain boundary and/or surface energies to the total energy. The higher stability of metastable phases in nanocrystals has been used to generate various metastable phases such as anatase-$TiO_2$ [44], tetragonal-$ZrO_2$ [45], and monoclinic-$Y_2O_3$ [46]. In this study, hydrothermal treatment eases overcoming the activation energy for the cubic-to-hexagonal phase transition in CdS at a low temperature, while pure low-temperature thermal treatment is not effective in inducing this phase transformation. As shown in Fig. S1, no phase transformation is detected when the powder is heated under similar conditions applied for the hydrothermal process (453 K and 20 h) but without NaOH addition, confirming the importance of NaOH in achieving the hexagonal phase. During the hydrothermal treatment, CdS is initially dissolved in NaOH and after saturation of the solution, CdS starts to precipitate again in the form of the thermodynamically stable hexagonal phase. By considering the reported hydrothermal process of CdS in the literature [47], the process for phase transformation in the alkaline solution can be written as follows.

$$CdS \text{ (Cubic)} \rightarrow Cd^{2+} + S^{2-} \qquad (1)$$

$$Cd^{2+} + 2(OH)^- \rightarrow Cd(OH)_2 \qquad (2)$$

$$Cd(OH)_2 + S^{2-} \rightarrow CdS \text{ (Hexagonal)} + 2(OH)^- \qquad (3)$$

These hydrothermal reactions are expected to be feasible in the presence of other alkaline solutions such as $Na_2S + H_2O$ [48,49].

Raman spectra of all samples are shown in Fig. 2b, indicating three important issues. First, all spectra exhibit three major peaks: the peak at 300 cm$^{-1}$ corresponding to the 1LO, the peak at 600 cm$^{-1}$ corresponding to its overtone 2LO, and the peak at 900 cm$^{-1}$ corresponding to 3LO (LO stands for longitudinal optical phonons) [50]. The ratios of the 2LO intensity to the 1LO intensity ($I_{2LO}/I_{1LO}$), which are presented in Table 1, should usually increase with the increase in particle size [51,52]. Therefore, the data in Table 1 indicate the enlargement of particle size of CdS after the HPT process with the $I_{2LO}/I_{1LO}$ ratio increasing from 0.28-0.30 to 0.34-0.46. Second, the Raman patterns can be effectively used to quantify the strain distribution by considering the Raman peak shifts [52]. Fig. 2c shows the Raman peak shifts to a low wavenumber after HPT processing, indicating the presence of residual tensile strain. When tensile stress is induced in a crystal, the distance between atoms increases, thus leading to the stretch of the chemical bond length [53]. With increasing the bonding length, the force constant is reduced, resulting in a decrease in the vibration frequency [54]. In this study, although CdS is subjected to HPT under a compressive stress of 6 GPa, it produces residual tensile strain due to the formation of lattice defects [33]. It was reported that tensile strain effectively regulates the electronic band structure to improve the charge transfer rate at interfaces and enhance the photocatalytic activity [55]. Third, the broadening of the Raman peak after HPT processing shown in Fig. 2c and given numerically in Table 1 indicates the formation of sulfur vacancies [51]. Vacancy-type defects can be accomplished as trap sites for electrons to ease carrier spatial separation [56].

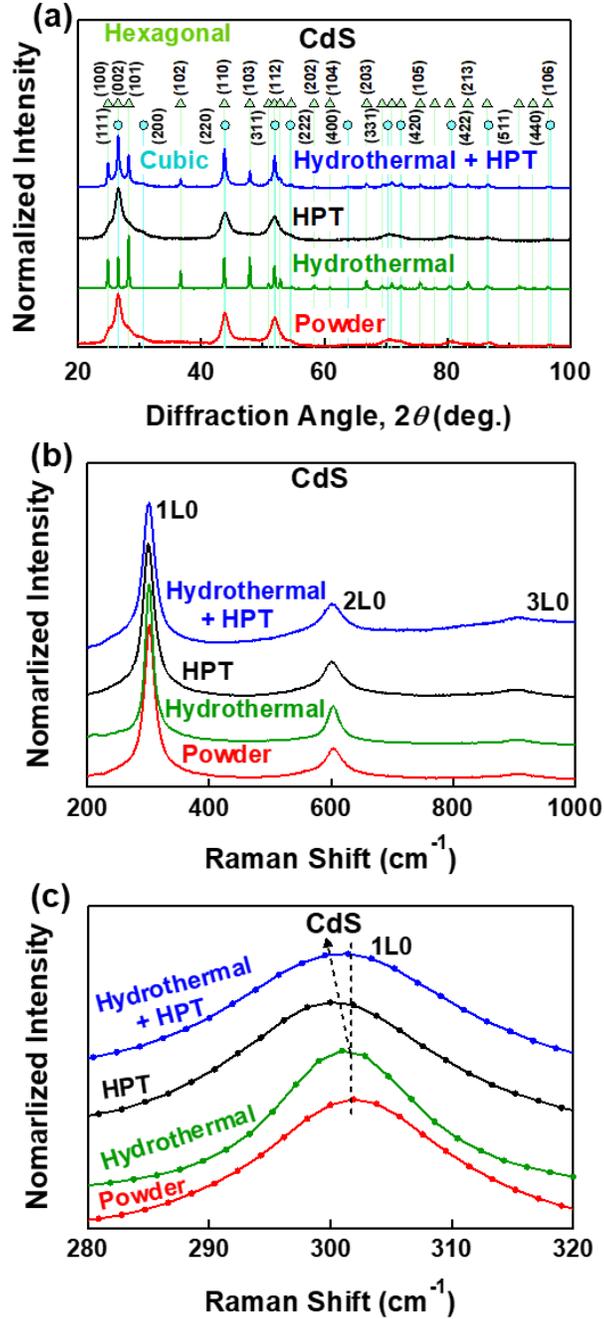

Figure 2. Phase transformation, tensile strain formation, and sulfur vacancy introduction in CdS sample. (a) XRD spectra and (b) Raman spectra and (c) magnified view of 1LO Raman peak for initial CdS powder, and hydrothermal-processed powder, HPT-processed powder, and hydrothermal+HPT-processed samples. Reference data for XRD profiles of cubic and hexagonal phases of CdS were taken from JCPDS card numbers 79-6256 and 89-2944, respectively.

Table 1. Phase fractions, crystallite size, Raman 2L0-to-1L0 peak intensity ratio, full width at half maximum (FWHM) of Raman 2L0 peak, and grain size estimated from TEM for initial CdS powder, and hydrothermal-processed, HPT-processed and hydrothermal+HPT-processed materials.

| Characteristics | Initial Powder | Hydrothermal | HPT | Hydrothermal+ HPT |
|---|---|---|---|---|
| Phase fractions (wt%) | Cubic: 46.3, Hexagonal: 53.7 | Cubic: 0.8, Hexagonal: 99.2 | Cubic: 97.4, Hexagonal: 2.6 | Cubic: 26.0, Hexagonal: 74.0 |
| Crystallite size (nm) | 3.2 | 40.7 | 2.0 | 2.5 |
| $I_{2LO}/I_{1LO}$ | 0.30 | 0.28 | 0.34 | 0.46 |
| FWHM of 2L0 ($cm^{-1}$) | 30 | 20 | 32 | 35 |
| Grain size (nm) | 25±6 | 73±20 | 9±5 | 12±5 |

The morphology of four samples, examined by SEM, and their nanostructures, examined by TEM with BF, SAED, and HR modes, are shown in Fig. 3. Examination of these micrographs indicates several major points. First, SEM indicates that the initial CdS powder is in the spherical form, but small nanorods are partly formed after the hydrothermal process. After HPT processing, the particle size increases to the range of tens of µm. The increase in the particle size is due to the effect of severe plastic deformation and high pressure from the HPT method [57]. The processing procedure also makes changes to the surface area of CdS. The specific surface areas examined by the BET method are 54.01, 11.90, 7.16, and 4.37 $m^2/g$ for the initial powder, hydrothermal sample, HPT sample, and hydrothermal+HPT sample, respectively. Second, the BF images and ring patterns of SAED verify the presence of numerous nanocrystals in all samples. The grain size measured from TEM are 25±6, 73±20, 9±5, and 12±5 nm for CdS initial powder, and hydrothermal, HPT and hydrothermal+HPT samples, respectively. Third, the formation of nanorods after the hydrothermal process can be observed while limited numbers of these nanorods can survive even after HPT processing. Fourth, the lattice images in Fig. 4f and 4g exhibit significant lattice distortion within nanocrystals, which is consistent with the Raman peak shift. Fifth, the presence of dislocations is evident after HPT processing, which can be considered a positive feature for this work because previous studies indicated that dislocations could amplify light absorption and improve photocatalytic activity [58]. Moreover, the residual stress from the presence of dislocations and vacancies can regulate interfacial electron transport [59]. Sixth, the interphase boundaries between cubic and hexagonal phases can be observed in all samples. These interphases can serve as heterojunctions to facilitate the separation and mobility of charge carriers to increase photocatalytic activity [60].

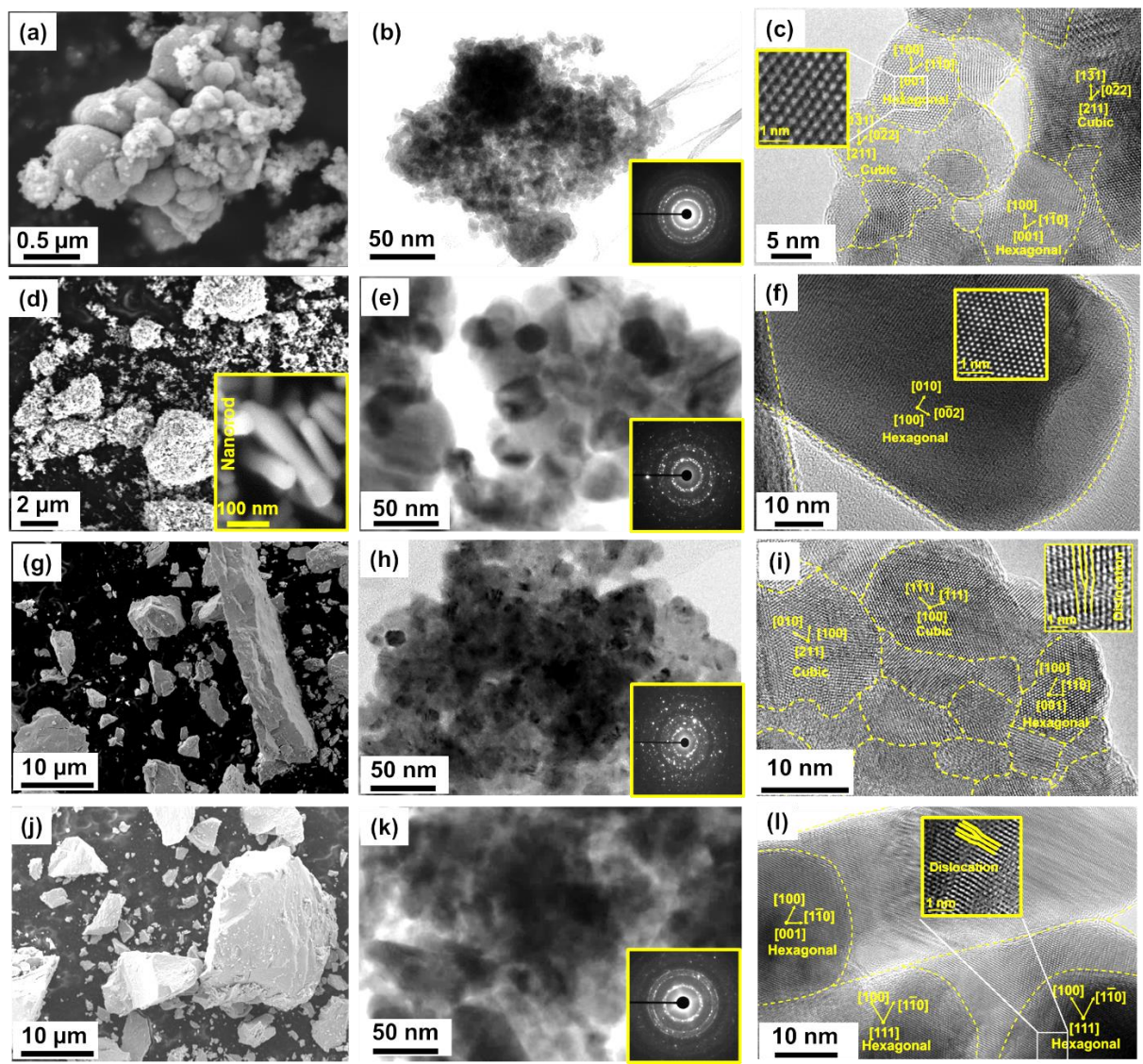

Figure 3. Coexistence of grain boundaries and dislocations in CdS after hydrothermal and HPT processes. (a,d,g,j) SEM images, (b,e,h,k) TEM BF images and SAED patterns and (c,f,i,l) TEM HR images for (a-c) initial CdS powder, and for sampled treated by (d-f) hydrothermal (g-i) HPT, and (j-l) hydrothermal+HPT.

A summary of structural and microstructural quantifications is given in Table 1, indicating that while particle size increases after HPT processing, the grain size and crystallite size decrease. It should be noted that the particle size, as an indicator of powder size, was measured by SEM; however, there can be multiple grains and crystallites in one particle. Crystallite size is the size of a single coherent domain within a particle which is measured by XRD, while grain size is the distance between high-angle grain boundaries which is measured by TEM. The resulting large particle size after HPT processing is due to the partial powder consolidation by high pressure and strain during HPT [33], while the small crystallite and grain sizes are due to the effect of plastic strain on microstructural refinement [57].

XPS spectra of four samples are shown in Fig. 4 for (a, b) Cd 3d and (c) S 2p, where (b) shows a magnified view of Cd $3d_{5/2}$ (all XPS peaks were adjusted by the C 1s peak at 248.8 eV). The valence states are $Cd^{2+}$ and $S^{2-}$ for all samples [61-63], but the peak shifts for Cd 3d after HPT processing, presented by the dotted lines in Fig. 4b, indicate the formation of sulfur vacancies [63]. The peak shift for S 2p is not visible in Fig. 4c, perhaps because of the close positions of peaks for sulfur anion and sulfur vacancies (about 1 eV difference) as well as because of the low fraction of sulfur vacancies [62-64]. To further examine the formation of vacancies, EPR measurement was performed as EPR is a powerful tool to detect vacancies in paramagnetic materials [64]. The EPR spectra for all catalysts are presented in Fig. 4d, indicating the low fraction of vacancies in power and hydrothermal samples while the peaks of sulfur vacancies appear after HPT processing at *g* = 2.003 [65]. Since EPR shows the signal for only sulfur vacancies that trap unpaired electrons such as one electron ($V_S•$), sulfur vacancies that trap two electrons ($V_S••$) or no electrons ($V_S$) cannot be detected by EPR measurement. Moreover, EPR peak intensity is dependent on the phase type, and thus, the EPR peak intensities for the two HPT-processed samples in Fig. 4d do not necessarily correspond to the number of vacancies. However, the appearance of EPR peaks after HPT processing is in line with the peak shifts in Raman and XPS spectra which suggests the formation of sulfur vacancies in the HPT-processed samples is in good agreement with earlier publications on the formation of nitrogen and oxygen vacancies [66,67]. The formation of vacancies can generally boost the photocatalytic activities of the materials [28,29,32]. The formation of vacancies by HPT, which is classified as a severe plastic deformation method, is due to the effect of plastic strain on the formation of lattice defects (i.e. strain-induced defects which are a source of work hardening phenomenon) [68,69]. Moreover, high pressure in HPT suppresses the mobility and annihilation of vacancies by affecting the vacancy migration energy [68,69], leading to supersaturated vacancy concentrations not only in metallic materials [57] but also in ceramics [33]. With further increases in the plastic strain, the material reacts to the deformation by transitioning from point defects (vacancies) to line defects (dislocations) and finally to planer defects (grain boundaries) [58].

Steady-state PL spectra of four samples are presented in Fig. 4e. The starting CdS powder has a broad peak at a wavelength of 611 nm which corresponds to the recombination of charge carriers on surface defects, while the hydrothermal sample possesses two PL peaks of band-edge emission at 506 nm wavelength and surface defect emission at a higher wavelength of 623 nm [1,2,8]. After HPT processing, no band-to-band emission is detected, and there are shifts in the wavelength of defect-related peaks. Some studies suggested that the enlargement of particle size and the introduction of point defects can lead to a peak shift which is consistent with this study [46,70,71]. Another study suggested that complex defects formed in AIIBVI nonstoichiometric compounds such as CdS can produce new states and significantly change PL peak positions [72]. A combination of particle size, point defects and complex defects together with the difference in the crystal structure should be responsible for peak shifts observed after HPT processing of the initial powder and the hydrothermal sample. Future studies need to be performed to identify the PL peak positions for different kinds of defects in different CdS phases. The intensity of peaks corresponding to defects decreases after HPT compared to samples before HPT processing, indicating that the synthesis by HPT process does not negatively affect the recombination of photo-generated holes and electrons, which is promising for CdS photocatalysts [18-22].

Fig. 4f presents the photocurrent spectra of four catalysts. The initial and HPT-processed CdS exhibit a steady photocurrent with a sharp decrease in photocurrent after turning off the light. The hydrothermal-processed CdS displays the spike peak at the start of light irradiation, followed by a rapid decrease to a steady state. This behavior suggests that while electron-hole pairs are initially separated, they quickly recombine, which can be also observed by the high PL intensity for this sample in Fig. 4d. However, by combining the hydrothermal and HPT processes, the intensity of these initial spike peak is significantly reduced and the drop in photocurrent over time becomes rather gradual, indicating that the combined hydrothermal and HPT process can effectively suppress electron-hole recombination leading to longer lifetime of photoelectrons. Here, it should be noted that because of the large particle sizes of samples and limitations in heating the samples (due to thermal annihilation of vacancies), it was hard to make dense films to compare the photocurrent data for the four samples quantitatively.

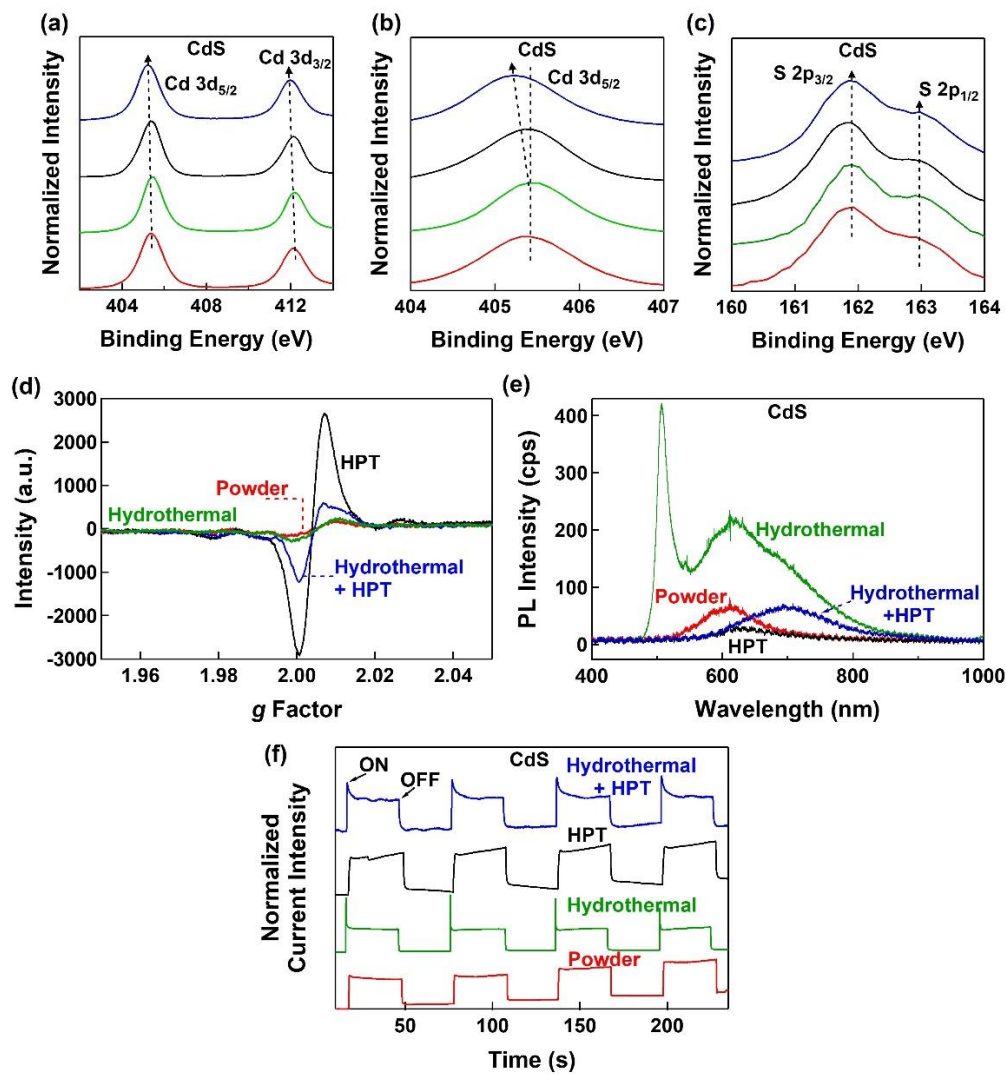

Figure 4. Formation of sulfur vacancies and suppression of radiative electron-hole recombination in CdS after hydrothermal and HPT processes. (a, b) XPS spectra of Cd 3d in two different magnifications, (c) XPS spectra of S 2p, (d) EPR spectra, (e) PL spectra, and (f) photocurrent generation spectra of initial CdS powder, and hydrothermal-processed, HPT-processed, and hydrothermal+HPT-processed samples.

The optical property and band structure of samples are exhibited in Fig. 5. The light absorbance of all samples in Fig. 5a is reasonably similar except for a lower light absorbance of the hydrothermal-processed sample. The bandgaps, calculated by Kubelka-Munk theory and Tacu plots in Fig. 5b are 1.94, 2.24, 1.96, and 1.92 eV for initial powder, hydrothermal, HPT, and hydrothermal+HPT samples, respectively. The hydrothermal sample possesses the widest optical bandgap due to the absence of defects. The XPS profiles used to determine the top of the valence band are presented in Fig. 5c, indicating that the valence band top energy slightly increases after HPT processing, due to the formation of defect states close to the valence band [55]. The conduction band bottom energy, determined by deducting the bandgap from the valence band top energy, was used for achieving the band structure as illustrated in Fig. 5d. All samples satisfy the energy requirement for hydrogen production and the formation of active species for photoreforming [73].

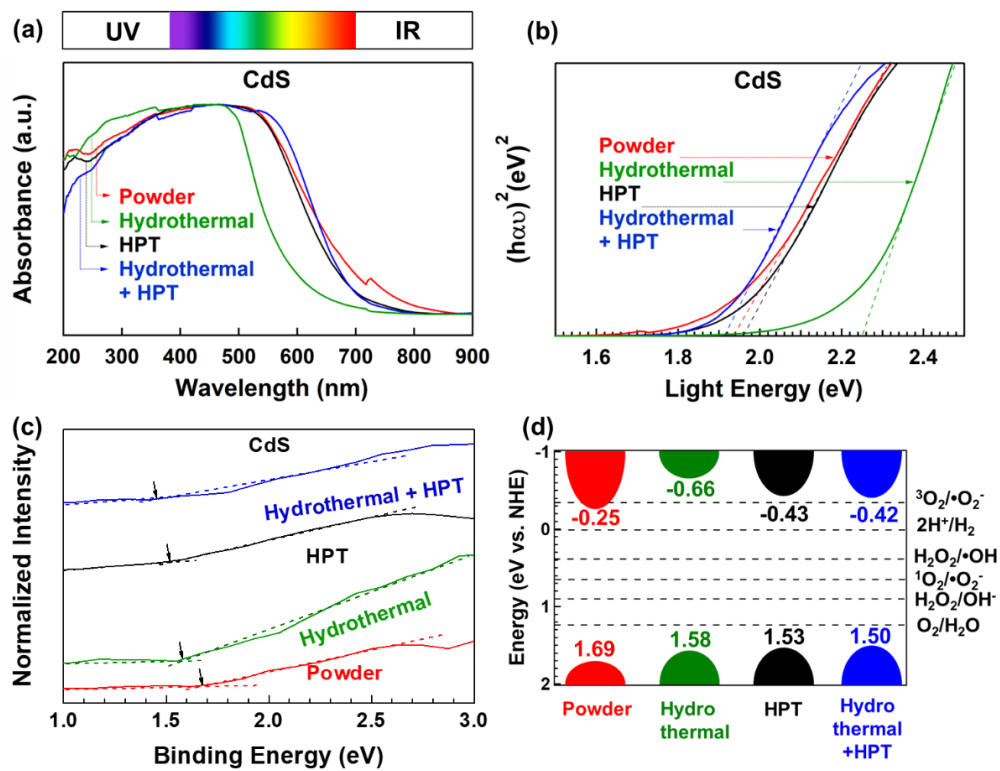

Figure 5. Suitable light absorption and band structure of CdS for photorefoforming. (a) UV-vis spectra, (b) calculation of bandgap via Kubelka-Munk theory ($\alpha$: photon absorbance, $h$: Planck's constant, $\nu$: photon frequency), (c) XPS of valence band top and (d) band structure in comparison with energy requirements for generation of hydrogen, oxygen and active species for initial CdS powder, and hydrothermal-processed, HPT-processed, and hydrothermal+HPT-processed samples.

## 3.2. Photocatalysis

During the photoreforming process, hydrogen production occurs alongside the degradation of PET plastic. The data demonstrate that all CdS catalysts exhibit photocatalytic activity for hydrogen production via PET photoreforming without the need for a co-catalyst, as shown in Fig. 6a. No hydrogen production is detected initially when the catalysts are stirred in the solution in the dark for 30 min. Moreover, in the blank test without catalyst addition, under light irradiation, no hydrogen is produced within 4 h. The amounts of hydrogen production are 3.6, 23.8, 35.4, and 83.4 µmol m$^{-2}$ for the initial powder, hydrothermal sample, HPT sample, and hydrothermal+HPT sample, respectively. Hydrogen production increases by HPT processing and a combination of hydrothermal and HPT methods results in 23 times higher efficiency compared to the initial CdS powder. The role of a co-catalyst in enhancing photoreforming activity, examined for the initial powder and the hydrothermal+HPT sample in Fig. 6b, indicates that the addition of 1 wt% platinum co-catalyst enhances the photocatalytic activity from 15.0 to 143.7 µmol m$^{-2}$. Specifically, hydrogen production in the hydrothermal+HPT CdS catalyst only doubles with the co-catalyst addition, suggesting that the co-catalyst can be avoided in this system by increasing the reaction time by a factor of only two. This improvement in photoreforming activity even in the absence of a co-catalyst suggests a promising dopant-free strategy for producing CdS photocatalysts with superior activity. Such improvement should be due to the stabilization of the hexagonal phase [19,27] as well as the generation of vacancies [28,29,32].

The liquid phase resulting from the photoreforming process was analyzed by $^1$H NMR to identify the degradation products of PET plastic. Fig. 7a shows the $^1$H NMR spectrum after 4 h of stirring without light irradiation, while Fig. 7b-f presents the $^1$H NMR spectra of four catalysts after 4 h of irradiation and photoreforming: (b) initial powder, (c) hydrothermal sample, (d) HPT sample, and (e, f) hydrothermal+HPT sample. For the sample stirred in NaOH under the dark condition, peaks of terephthalate and ethylene glycol are visible which is consistent with the expected products from the hydrolysis of PET in NaOH. It is known that PET hydrolysis in NaOH involves the formation of its monomer terephthalate, which is chemically stable and reused in PET production, and ethylene glycol, which is further oxidized into various useful compounds by photoreforming [39,73-76]. The photoreforming products identified by NMR include formic acid for the CdS initial powder, formic acid and acetic acid for the hydrothermal sample, formic acid and acetic acid for the HPT sample, and formic acid, glyoxal, glycolic acid, and acetic acid for hydrothermal+HPT sample. The PET conversion pathway is depicted in Fig. 7e by considering NMR results and information reported in previous studies [74-77]. Earlier studies also reported that the photoreforming products depend on the catalyst features. For instance, $MoS_2/Cd_xZn_{1-x}S$ catalyst produces formic acid, glycolic acid and methylglyoxal from PET photoreforming [74], carbonized polymer dots - $C_3N_4$ catalyst yields glycolic acid, glycolaldehyde and ethanol [76], and using a $C_3N_4$ - carbon nanotubes - NiMo hybrid catalyst leads to glyoxal and glycolic acid formation [75].

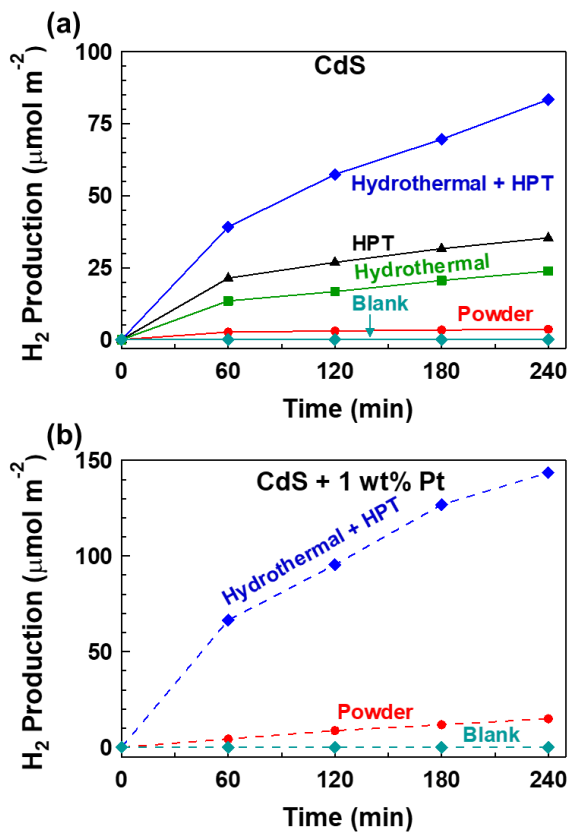

Figure 6. Enhancement of hydrogen production from photocatalytic PET plastic degradation by phase transformation and sulfur vacancy introduction in CdS. Hydrogen production versus irradiation time during photocatalytic process with PET plastic as sacrificial agent (a) without co-catalyst addition and (b) with 1wt% platinum addition for initial CdS powder, and hydrothermal-processed, HPT-processed and hydrothermal+HPT-processed samples.

Fig. 9h presents the quantitative $^1H$ NMR data using maleic acid as an internal standard. For initial CdS power, 0.19 mmol m$^{-2}$ formic acid is obtained. In the photoreforming system using hydrothermal powder, 0.30 mmol m$^{-2}$ formic acid, 0.14 mmol m$^{-2}$ acetic acid, and 0.02 mmol m$^{-2}$ methanol are formed. Compared to the initial powder, the HPT process also leads to an increase in the amount of photodegradation products with 0.24 mmol m$^{-2}$ formic acid and 0.16 mmol m$^{-2}$ acetic acid. In the system of the hydrothermal+HPT sample, the degradation products are formic acid, acetic acid, glyoxal, and glycolic acid with the amount of 0.39 mmol m$^{-2}$, 0.06 mmol m$^{-2}$, 1.97 mmol m$^{-2}$, and 1.18 mmol m$^{-2}$, respectively. The data confirm that the integration of hydrothermal and HPT processes enhances photocatalytic activity for both hydrogen production and plastic conversion. Here, it should be noted that the structures of catalysts remain unchanged after photoreforming both in bulk (analyzed by XRD) and on the surface (analyzed by Raman spectroscopy), as shown in Fig. S2.

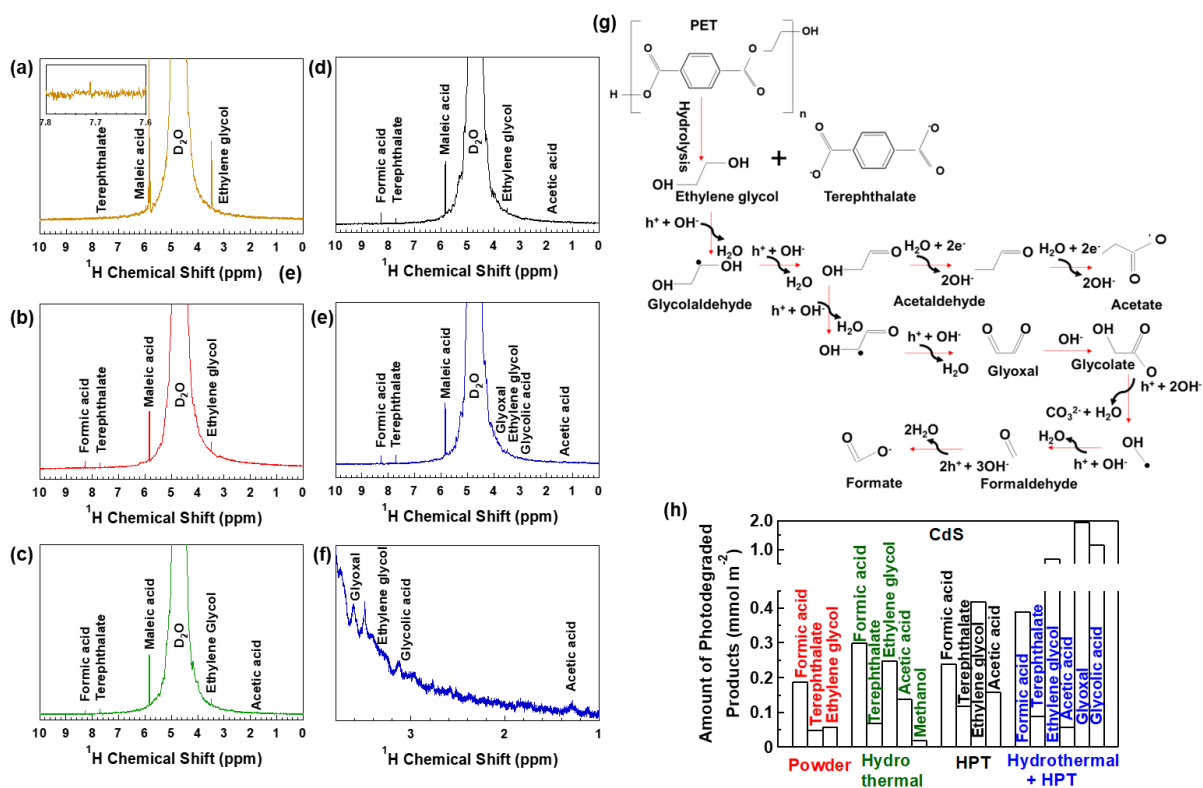

Figure 7. Photodegradation of PET plastic by CdS catalysts. $^1$H NMR spectra of degraded PET in (a) dark condition and (b-f) after 4 h light irradiation for (b) initial CdS powder, and (c) hydrothermal-processed, (d) HPT-processed, (e,f) hydrothermal+HPT-processed samples. (g) Proposed degradation pathway for PET. (h) Quantitative $^1$H NMR analysis for photodegraded products from PET plastic after 4 h light irradiation.

## 3.3. Computational results

A comparative study between the two CdS phases was performed using first-principles calculations. As the DFT tends to underestimate the bandgap calculations, a correction using DFT+$U$ was performed. The crystallographic information of the modeled structures in comparison with experimental data is summarized in Table 2, indicating good agreement between the calculations and the experimental results.

The bandgap obtained for different $U$ values is shown in Fig. 8. By increasing the $U$ value, the bandgap values for both phases gradually increase. However, the trend suggests that the hexagonal phase has a smaller bandgap than the cubic phase for all selected $U$ values. This finding is in good agreement with earlier studies [77-79]. Since the experimental bandgap of CdS was reported in the range of 2.1-2.4 eV for the two phases, $U$ values in the range of 8-11 can predict the experimental values. Here, for the rest of the calculations, a $U$ value of 9 eV was considered as a number that can reasonably cover the whole experimental data.

Table 2. Crystallographic information of pristine CdS cubic and hexagonal structures obtained by DFT calculations after optimization compared with experimental values.

| Structure | Space Group | Method | Angles (°) | | | Lattice Parameters (Å) | | | ICDD Number |
|---|---|---|---|---|---|---|---|---|---|
| | | | $\alpha$ | $\beta$ | $\gamma$ | $a$ | $b$ | $c$ | |
| Cubic | $F$-43$m$ | Experimental | 90 | 90 | 90 | 5.83 | 5.83 | 5.83 | 1011054 |
| | | Calculation | 90 | 90 | 90 | 5.93 | 5.93 | 5.93 | |
| Hexagonal | $P6_3mc$ | Experimental | 90 | 90 | 120 | 4.13 | 4.13 | 6.71 | 1011260 |
| | | Calculation | 90 | 90 | 120 | 4.20 | 4.20 | 6.84 | |

To understand the effect of oxygen vacancies, a supercell consisting of 64 atoms was built for each phase and 3 and 6 at% of vacancies were introduced. Fig. 9 shows the calculated DOS for the cubic (Fig. 9a, 9c and 9e) and the hexagonal (Fig. 9b, 9d and 9f) phases with vacancy fractions of (a, b) 0, (c, d) 3 and (e, f) 6 at%. For both cases, the calculated DOS plots show that the valence band consists mainly of S 3$p$ states, while the conduction band consists mainly of Cd 5$s$ states [77,80]. In addition, the DOS shows that bandgap decreases with increasing the concentration of sulfur vacancies, but for the three studied states of vacancies, the hexagonal phase always exhibits a lower bandgap than the cubic phase. It should be noted that the vacancy formation energy for the hexagonal and cubic phases was calculated as 3.60-3.64 eV and 3.59-3.63 eV which are quite similar.

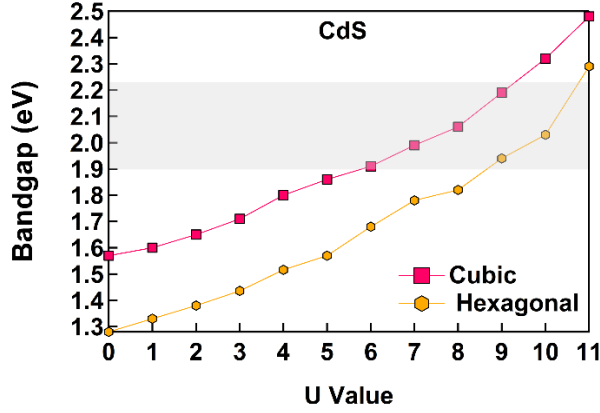

Fig. 8. Smaller bandgap of hexagonal phase compared to cubic phase in CdS. Calculated bandgap of cubic and hexagonal phase obtained by varying the Hubbard U-term for Cd *d* orbital in comparison with reported range of experimental data.

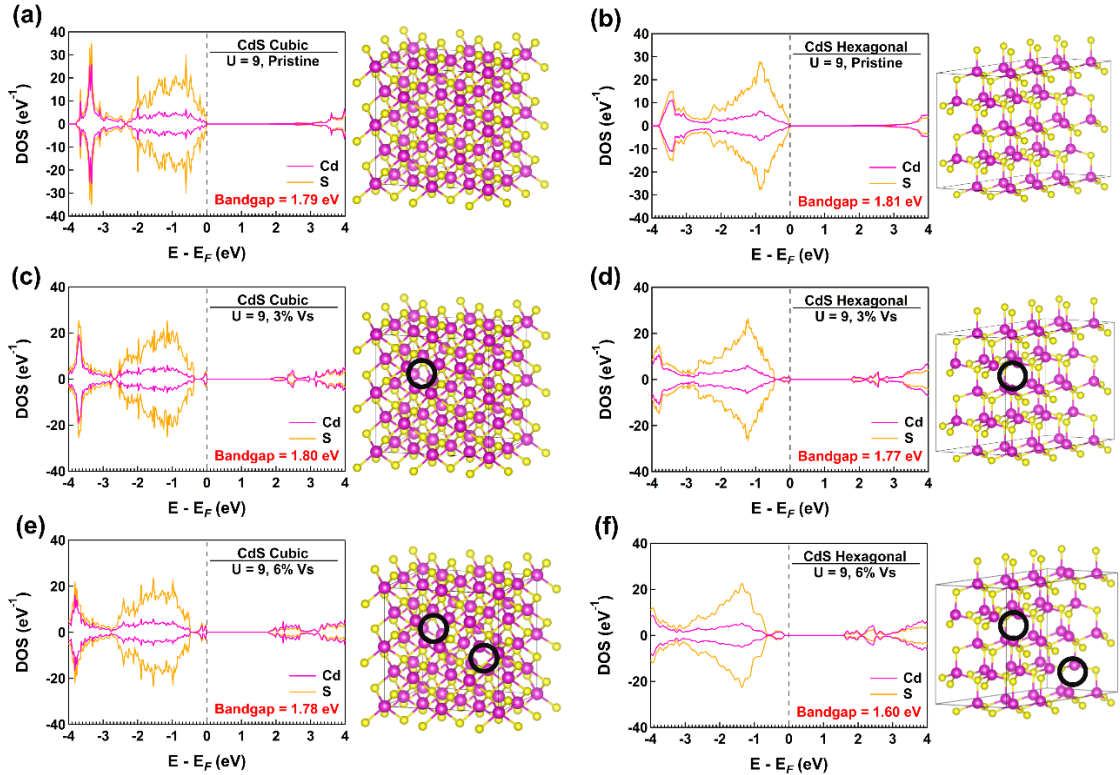

Figure 9. Reduction of bandgap of CdS by formation of sulfur vacancies. Calculated DOS with Hubbard *U* value of 9 eV and corresponding supercells for (a) 0%, (c) 3% and (e) 6% sulfur vacancies in cubic phase, and for (b) 0%, (d) 3% and (f) 6% sulfur vacancies in hexagonal phase. Black circles in supercells show position of sulfur vacancies.

The analysis of charge distribution around sulfur vacancies (Fig. S3) shows charge accumulation around the Cd atoms in the vicinity of vacancies. This charge redistribution provides electrostatic charges that lead to local lattice distortion [81]. The average bond length for both phases after optimization reaches 2.53 Å; however, the bond lengths around the vacancy reach 2.58 Å and 2.59 Å for the cubic and hexagonal phases, respectively. The differences in the bond length suggest a slightly higher strain in the hexagonal phase compared to the cubic phase. The distortion generated in the $CdS_4$ tetrahedra around the vacancies can lead to the formation of dipole moments and internal electric fields [77], facilitating the efficient photo-generated electron-hole separation, particularly in the hexagonal phase of CdS [20]. Hence, the presence of sulfur vacancies and resultant charge redistribution and lattice strain can provide active sites for reactant molecule adsorption and charge separation [77,81].

## 4. Discussion

This study reports the first application of the combined process of hydrothermal and HPT to control metastable cubic to stable hexagonal phase transformation in CdS and produce an active photocatalyst with abundant sulfur vacancies. The synthesized CdS shows high activity in simultaneous hydrogen production and PET plastic degradation to various valuable compounds such as formic acid, acetic acid, glyoxal, and glycolic acid even without co-catalyst addition. Two key issues should be further discussed: (i) the mechanisms for simultaneous hydrogen production and plastic degradation and (ii) reasons for the superior photocatalytic activity of synthesized CdS compared to commercial CdS powder.

Regarding the first issue, the mechanism of photoreforming for hydrogen production and plastic degradation resembles the traditional photocatalytic hydrogen production with sacrificial reagents. During the photoreforming process, light irradiation excites the CdS semiconductor, causing electrons to move from the valence band top to the conduction band bottom. These electron species then reduce $H^+$ in $H_2O$ to form $H_2$, while the holes left in the valence band either directly degrade the plastic or generate radicals such as •OH which subsequently oxidize the plastic into smaller organic molecules. Although quantification of hole fractions is challenging in this study, the fraction of •OH radicals can be experimentally quantified. As •OH radicals react with coumarin to form 7-hydroxycoumarin , coumarin was added to the photocatalytic system and 7-hydroxycoumarin formation through the following relation was evaluated by fluorescence as an indication of •OH radicals, as shown in Fig. 10a.

•OH + Coumarin → 7-Hydroxycoumarin (4)

The data in Fig. 10a show that the hydrothermal+HPT sample possesses the highest fluorescence intensity and thus the largest amount of •OH radicals. The reactions involved in the PET photoreforming process by CdS can be summarized below.

$$CdS + h\nu \rightarrow h^+_{(VB)} + e^-_{(CB)} \quad (5)$$

$$2H^+ + 2e^- \rightarrow H_2 \quad (6)$$

$$h^+_{(VB)} + OH^- \rightarrow \bullet OH \quad (7)$$

$$h^+_{(VB)}/\bullet OH + PET \rightarrow \text{Small Organic Products} \quad (8)$$

These reactions indicate the significance of •OH radicals and holes on photodegradation. Therefore, in the photoreforming process, one justification for the superior activity of the hydrothermal+HPT sample is the high quantity of reactive •OH species. To better understand the roles of reactive species, quenching experiments were performed for the hydrothermal+HPT sample and plastic degradation products and hydrogen amount were analyzed, as shown in Fig. 10b and 10c, respectively (see Fig. S4 for relevant NMR spectra). Adding both CH$_3$OH (scavenger for holes) as well IPA (scavenger for •OH radicals) leads to a reduction in plastic photodegradation products (Fig. 10b) and hydrogen production (Fig. 10c). Data indicate that the addition of CH$_3$OH results in a significantly greater reduction in the number of photogenerated products compared to the addition of IPA. It should be noted that the addition of EDTA-2Na as a hole scavenger led to similar results as CH$_3$OH. These findings suggest that holes play a more significant role than •OH radicals in the plastic degradation process, consistent with previously reported data [82].

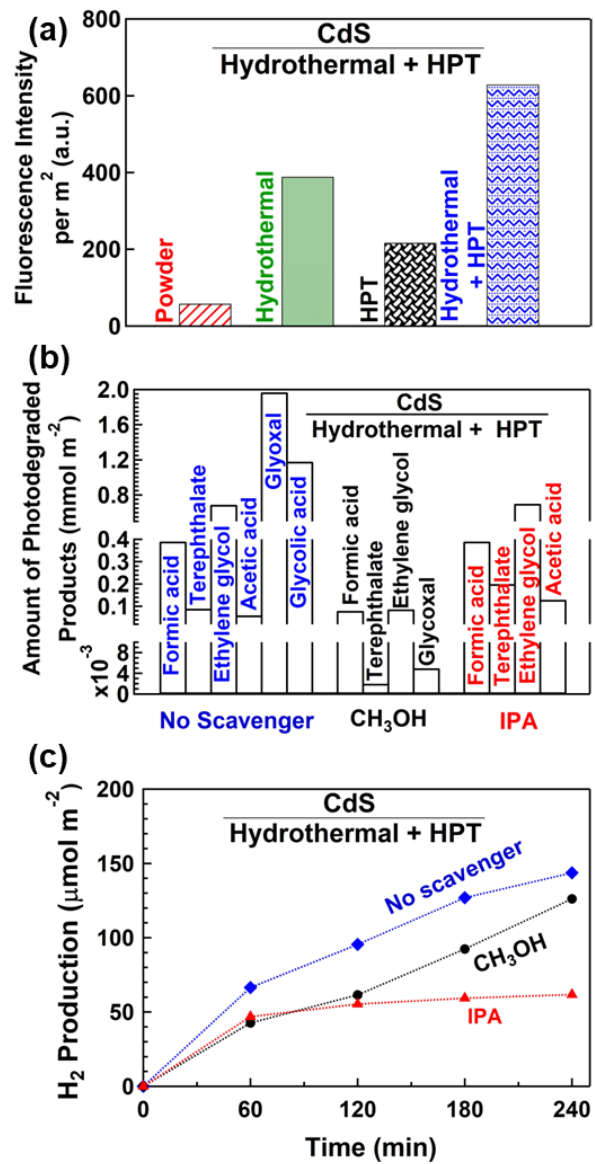

Figure 10. Enhancement of PET plastic degradation by phase transformation and sulfur vacancy introduction in CdS by contribution of holes and hydroxyl radicals. (a) Fluorescence intensity at 450 nm wavelength after light illumination for 20 min after adding coumarin as trapper of •OH radicals for initial CdS powder, and hydrothermal-processed, HPT-processed and hydrothermal+HPT-processed samples. (b) Quantitative $^1$H NMR analysis for photodegraded products from PET plastic after 4 h light irradiation and (c) hydrogen production versus irradiation time during photocatalytic process with and without adding $CH_3OH$ (scavenger of holes) and IPA (scavenger of •OH radicals) for hydrothermal+HPT-processed sample.

Regarding the second issue, it should be noted that the higher activity of CdS after hydrothermal and HPT treatment compared with the other three samples cannot be explained by only phase fractions. There are several factors contributing to the enhanced photoreforming performance of the hydrothermal+HPT sample including the presence of the majority of hexagonal phase, sulfur vacancies and hexagonal-cubic heterojunctions. Firstly, it is widely recognized that the hexagonal phase of CdS exhibits favorable optoelectronic properties and shows effective separation of photo-induced electrons and holes [27]. DFT calculations in this study confirm that the hexagonal phase shows a narrower bandgap than the cubic phase. With the introduction of vacancies, the hexagonal phase exhibits further bandgap narrowing, and slightly greater lattice strain compared to the cubic phase which can result in dipole moment formation and internal electric fields which are beneficial for electron-hole separation. The process used in this study increases the hexagonal phase fraction. Secondly, the formation of sulfur vacancies, confirmed by XPS on the surface, by Raman spectroscopy in the subsurface, and by EPR in bulk, can play a crucial role in altering CdS properties and enhancing photocatalytic performance. Although compared to oxygen vacancies, there are much fewer studies about sulfur vacancies [29,32], these vacancies are expected to enhance the light absorbance and act as activation sites for photocatalysis [28,29]. Some studies reported that sulfur vacancies provide a number of advantages, including extending the spectrum of photon absorption, modifying the electronic structure, promoting charge carrier movement and separation, and boosting surface reactions [83,84]. Excess electrons from the removal of sulfur can migrate to the empty orbitals of metals, causing shallow donor formation below the conduction band and increasing light absorption [35]. Thirdly, phase boundaries between cubic and hexagonal phases can act as heterojunctions that facilitate charge carrier migration. It was shown that the combination of oxygen vacancies and heterojunctions is more effective for charge carrier separation, and thus, the presence of sulfur vacancies and heterojunction is expected to be more beneficial [26,29,32]. Finally, other defects like dislocations and grain boundaries observed in TEM images can modulate interfacial electron transport, enhance light absorption, and promote the photoreforming process [58,59].

It is worthwhile to compare the catalytic activities achieved in this study with those reported in the literature, although such comparisons should be made cautiously due to variations in experimental conditions, including reactor type, irradiation source type and intensity, temperature, catalyst concentration, plastic concentration and alkaline substance concentration [85,86]. Table 3 presents the hydrogen production of synthesized CdS samples in this study in comparison with other reported data in the literature [15,75,77,82,87-90]. The catalyst after hydrothermal and HPT processing exhibits one of the highest activities reported for CdS-based catalysts, while, unlike many other reported catalysts, this catalyst is in the pure form. The introduction of sulfur vacancies in the stable phase of CdS without the addition of impurity atoms is in contrast to previous chemical procedures or doping-based methods [26]. Nevertheless, the benefits of using impurity atoms to produce sulfur vacancies may be undermined by impurity-induced recombination because such recombination was reported when oxygen vacancies are introduced by impurity atoms [83,84]. It is expected that vacancies with no effect of impurity generate active sites with a variety of coordination numbers and dangling bond properties more effectively [35]. Another difference between this study and other works is that the method used here generates vacancies in both the bulk and surface of the material [91,92], and some studies mentioned that the photocatalytic activity is enhanced by the co-presence of bulk and surface vacancies [93].

Table 3. Comparison of hydrogen production from PET photoreforming using pure CdS catalysts in this study with reported data of various catalysts.

| Catalyst | Produced $H_2$ ($\mu$mol m$^{-2}$ h$^{-1}$) | Ref. |
|---|---|---|
| CdS | 16.23 | [15] |
| Defect-rich NiPS$_3$ | 540.78 | [15] |
| CdS/Defect-rich NiPS$_3$ | 926.48 | [15] |
| MXene/Zn$_x$C$_{1-x}$S | 123 | [77] |
| Cd$_{0.5}$Zn$_{0.5}$S | 7.49 | [87] |
| Co$_x$O$_y$/Cd$_{0.5}$Zn$_{0.5}$S | 8.87 | [87] |
| RP@Co$_x$O$_y$/Cd$_{0.5}$Zn$_{0.5}$S | 59.27 | [87] |
| O-CuIn$_5$S$_8$ nanosheet | 27.39 | [88] |
| BiVO$_4$ | 140 | [89] |
| BiVO$_4$/0.05MoO$_x$ | 360 | [89] |
| BiVO$_4$/0.10MoO$_x$ | 110 | [89] |
| BiVO$_4$/0.15MoO$_x$ | 690 | [89] |
| CN$_{0.14}$ porous microtube | 0.42 | [90] |
| CN-CNTs-NiMo | 0.77 | [75] |
| Pt/g-C$_3$N$_4$ | 102 | [82] |
| Commercial CdS | 3.61 | This study |
| Hydrothermal CdS | 23.81 | This study |
| HPT CdS | 35.37 | This study |
| Hydrothermal+HPT CdS | 83.41 | This study |
| Commercial CdS/Pt | 15 | This study |
| Hydrothermal+HPT CdS/Pt | 143.67 | This study |

## 5. Conclusions

A two-step process of hydrothermal and HPT was used to introduce metastable-to-stable phase transformation and sulfur vacancies into pure CdS catalyst without impurity addition. The synthesized CdS catalyst was successfully utilized for simultaneous hydrogen production and PET plastic conversion without co-catalyst addition with a 23-fold higher activity than initial commercial CdS. Detailed analyses confirmed that PET plastic was converted into various useful organic compounds like formic acid, glyoxal, glycolic acid, and acetic acid mainly due to the contribution of holes and hydroxyl radicals. This experimental study which was combined with first-principles calculations initiated a new approach for phase transformation and vacancy engineering of sulfide catalysts to achieve simultaneous hydrogen production and plastic degradation.

## Acknowledgments

The author JHJ acknowledges a scholarship from the Q-Energy Innovator Fellowship of Kyushu University. This work was funded in part by Mitsui Chemicals, Inc., Japan, in part via a Grant-in-Aid from the Japan Society for the Promotion of Science (JSPS) (No. JP22K18737), in part by Japan Science and Technology Agency (JST), the Establishment of University Fellowships Towards the Creation of Science Technology Innovation (JPMJFS2132), in part by the ASPIRE project of JST (JPMJAP2332), in part by the University of Rouen Normandy through project BQRI MaP-StHy202, and in part by the CNRS Federation IRMA-FR 3095.

*Supporting Information*

**Phase and sulfur vacancy engineering in cadmium sulfide for boosting hydrogen production from catalytic plastic waste photoconversion**

Thanh Tam Nguyen[a,b], Jacqueline Hidalgo-Jiménez[a,c], Xavier Sauvage[d], Katsuhiko Saito[e], Qixin Guo[e], and Kaveh Edalati[a,b,]*

[a] WPI, International Institute for Carbon Neutral Energy Research (WPI-I2CNER), Kyushu University, Fukuoka 819-0395, Japan

[b] Mitsui Chemicals, Inc. - Carbon Neutral Research Center (MCI-CNRC), Kyushu University, Fukuoka 819-0395, Japan

[c] Graduate School of Integrated Frontier Sciences, Department of Automotive Science, Kyushu University, Fukuoka 819-0395, Japan

[d] Univ Rouen Normandie, INSA Rouen Normandie, CNRS, Groupe de Physique des Matériaux, UMR6634, 76000 Rouen, France

[e] Department of Electrical and Electronic Engineering, Synchrotron Light Application Center, Saga University, Saga 840-8502, Japan

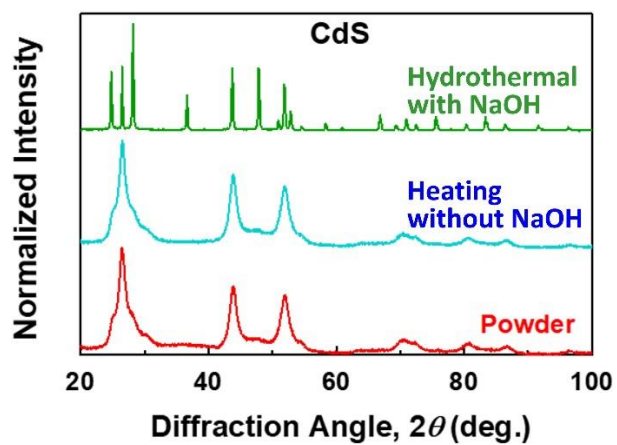

Figure S1. XRD profiles of initial CdS powder before heat treatment and after heating without NaOH addition at 453 K for 20 h and after hydrothermal treatment with NaOH addition at 453 K for 20 h.

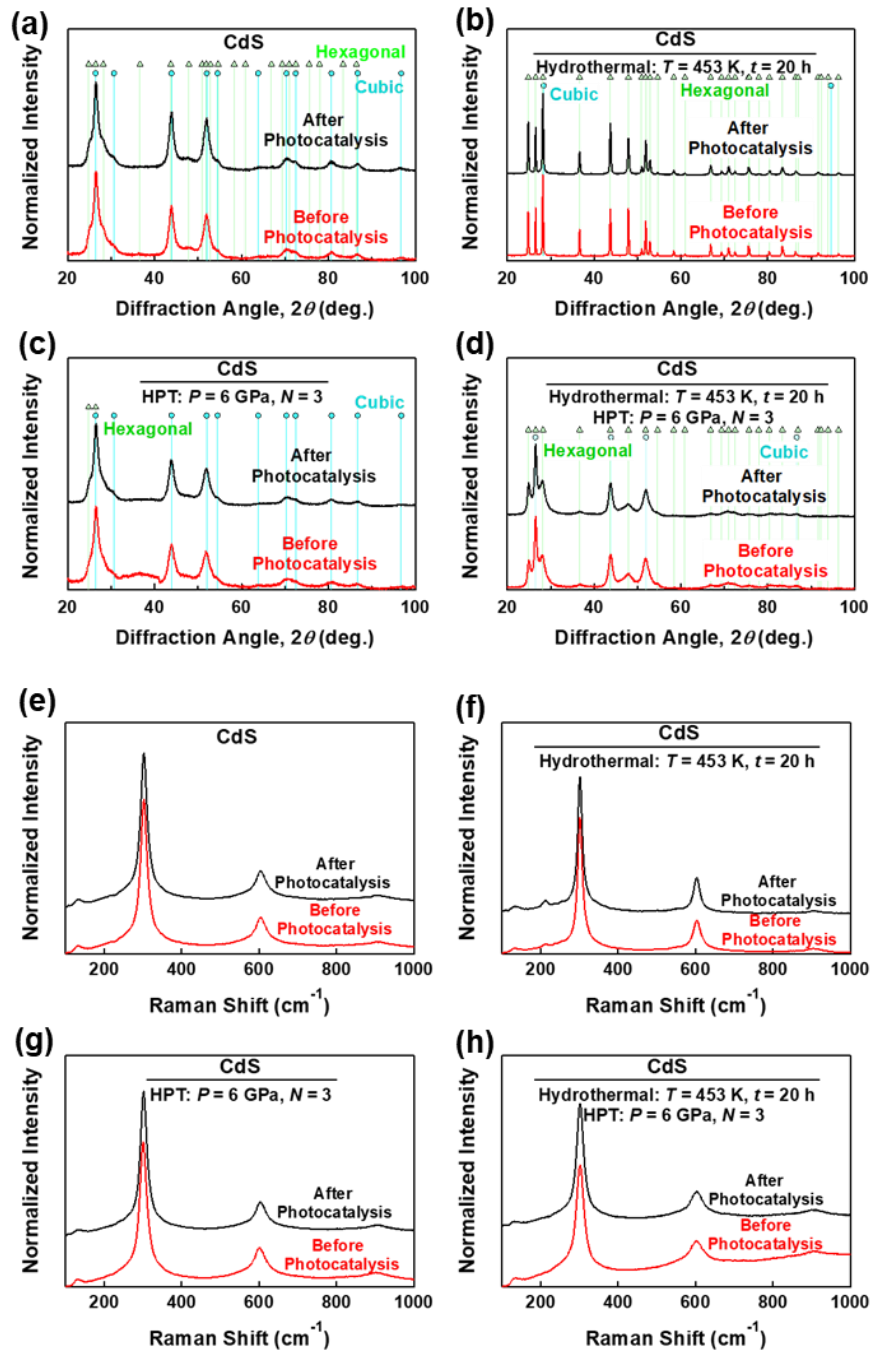

Figure S2. Stability of CdS catalysts for simultaneous photocatalytic hydrogen production and PET plastic degradation. (a-d) XRD and (e-h) Raman spectra for initial CdS powder, and hydrothermal-processed, HPT-processed, and hydrothermal+HPT-processed samples before and after photocatalytic test for 4 h.

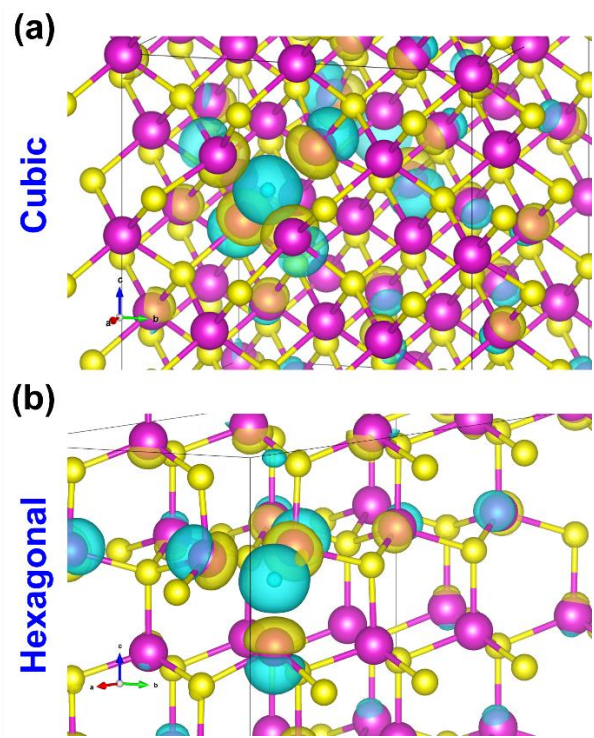

Figure S3. Charge density distribution on defective CdS with oxygen vacancies: (a) cubic phase, and (b) hexagonal phase. Yellow bubbles show electron accumulation and blue bubbles show electron depletion.

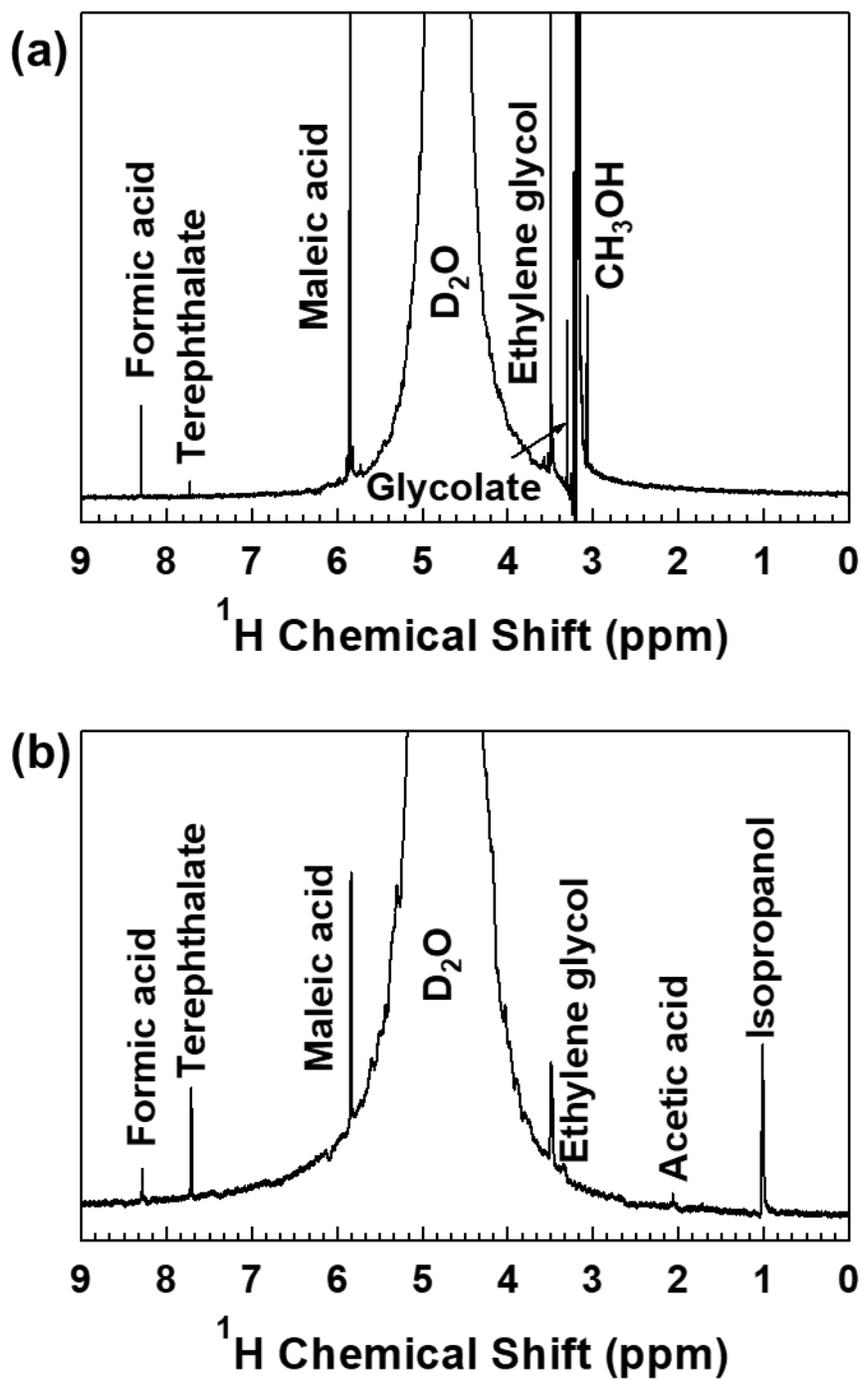

Figure S4. $^1$H NMR spectra of PET degradation products in 10 M NaOD with maleic acid as internal standard using (a) CH$_3$OH as hole scavenger and (b) Isopropanol as •OH radical scavenger.

E